\documentclass{article}
\usepackage{spconf,amsmath,graphicx,booktabs,makecell,stfloats}

\begin{document}
\title{THE MULTIMODAL INFORMATION BASED SPEECH PROCESSING (MISP) 2022 CHALLENGE: AUDIO-VISUAL DIARIZATION AND RECOGNITION}
%

\name{
\begin{tabular}{c}
Zhe Wang\textsuperscript{1}, Shilong Wu\textsuperscript{1}, Hang Chen\textsuperscript{1}, Mao-Kui He\textsuperscript{1}, Jun Du\textsuperscript{1,*}\thanks{\textsuperscript{*}corresponding author}, Chin-Hui Lee\textsuperscript{2}, \\ \textit{Jingdong Chen\textsuperscript{6}},  \textit{Shinji Watanabe\textsuperscript{3}}, \textit{Sabato Siniscalchi\textsuperscript{2,4}}, \textit{Odette Scharenborg\textsuperscript{7}}, \\ \textit{Diyuan Liu\textsuperscript{5}}, \textit{Baocai Yin\textsuperscript{5}}, \textit{Jia Pan\textsuperscript{5}}, \textit{Jianqing Gao\textsuperscript{5}}, \textit{Cong Liu\textsuperscript{5}}
\end{tabular}
} 
\address{
 \textsuperscript{1} University of Science and Technology of China, China  \textsuperscript{2} Georgia Institute of Technology, USA\\
 \textsuperscript{3} Carnegie Mellon University, USA \textsuperscript{4} Kore University of Enna, Italy \textsuperscript{5} iFlytek, China\\ 
 \textsuperscript{6} Northwestern Polytechnical University, China \textsuperscript{7} Delft University of Technology, The Netherlands
}
\ninept
\maketitle
\begin{abstract}
The Multi-modal Information based Speech Processing (MISP) challenge aims to extend the application of signal processing technology in specific scenarios by promoting the research into wake-up words, speaker diarization, speech recognition, and other technologies. The MISP2022 challenge has two tracks: 1) audio-visual speaker diarization (AVSD), aiming to solve ``who spoken when'' using both audio and visual data; 2) a novel audio-visual diarization and recognition (AVDR) task that focuses on addressing ``who spoken what when'' with audio-visual speaker diarization results. Both tracks focus on the Chinese language, and use far-field audio and video in real home-tv scenarios: 2-6 people communicating each other with TV noise in the background. This paper introduces the dataset, track settings, and baselines of the MISP2022 challenge. Our analyses of experiments and examples indicate the good performance of AVDR baseline system, and the potential difficulties in this challenge due to, e.g., the far-field video quality, the presence of TV noise in the background, and the indistinguishable speakers.

\end{abstract}
\section{Introduction}
\label{sec:intro}
Modern speech-enabled systems still suffer from performance degradation in real-world scenarios (e.g., at home and in meetings) due to factors associated with adverse acoustic environments and conversational multi-speaker interactions. Inspired by the finding that visual cues can help human speech perception \cite{rosenblum2008speech}, many researchers have proposed to use the visual modality to improve acoustic robustness \cite{afouras2018deep,son2017lip}. The MISP2021 challenge \cite{MISP} released a large distant multi-microphone conversational Chinese audio-visual corpus, and some advanced audio-visual speech recognition (AVSR) systems have been proposed \cite{xu2022channel,wang2022sjtu}. However, these systems assume that the correspondence between speech segments and speakers is known in advance, which greatly limits its scope in real-world applications. For the second MISP challenge, we target the problem of audio-visual speaker diarization (AVSD), and audio-visual diarization and recognition (AVDR) in the home-tv scenarios. Specifically, the AVDR is an extended task from AVSR, replacing oracle speaker diarization results with AVSD results.

Many approaches have been proposed on speaker diarization and speech recognition under the audio-only condition. \cite{chime6-xvector} utilized x-vector \cite{xvector}, agglomerative hierarchical clustering (AHC) and an LSTM-based overlap detector to get diarization results, which can be used for the guided source separation (GSS) and the deep neural network-hidden markov model (DNN-HMM). \cite{chime6-tsvad} proposed a novel system, which includes a speaker diarization module with target-speaker voice activity detection (TS-VAD), and a speech recognition module with self-attention. However, the audio-only speaker diarization and speech recognition task in the real scenes is still a huge challenge because of the potential strong background noise and high ratios of overlapping speech \cite{dihard}.

Facial behavior is highly correlated with speech activity \cite{facial}, and visual modality is not disturbed by harsh acoustic environment. Researchers show great interest in audio-visual speaker diarization (AVSD) and audio-visual speech recognition (AVSR). For AVSD system, some related works have been proposed. \cite{avsd2} utilized mutual information to fuse the audio and video modalities, while \cite{avsd3} used a Bayesian method for audio-visual speaker diarization. In recent years, many deep learning methods have emerged. \cite{avsd4} used an audio-visual synchronization model, \cite{avsd5} proposed a diarization method using self-supervised learning, achieving positive results. For AVSR system, \cite{son2017lip} proposed a ‘Watch, Listen, Attend and Spell’ (WLAS) network on the LRS data set and \cite{afouras2018deep} adopted a Transformer-based model. \cite{ma2021end} developed a CTC/Attention model based on conformer blocks. A DNN-HMM hybrid AVSR system with a gating layer \cite{tao2018gating} also showed good performance. Although AVSD and AVSR have received increased attention and have been shown to significantly outperform conventional audio-only methods, there is as yet little research done on audio-visual diarization and recognition (AVDR) which concentrates on AVSR with AVSD results.

The MISP2022 challenge includes two tracks: audio-visual speaker diarization (AVSD), and audio-visual diarization and recognition (AVDR). In this paper, we discuss the MISP2022 challenge, the data, tracks, and provide a detailed description of the baseline AVDR system, followed by a deep analysis. Besides, we point out the difficulties that participants may encounter in this challenge, including the low quality of far-field videos, the background noise in the home-tv scenarios, and the existence of indistinguishable speakers. To the best of our knowledge, we proposed a brand-new AVDR task, and our proposed AVDR baseline system is the first to concatenate the AVSD and AVSR into one large system. The resulting system has broad application prospects. More challenge details\footnote{https://mispchallenge.github.io/mispchallenge2022} and the baseline code\footnote{https://github.com/mispchallenge/misp2022\_baseline} can be found on the websites.


\begin{figure*}[htbp]
\centering
\includegraphics[width=1\textwidth]{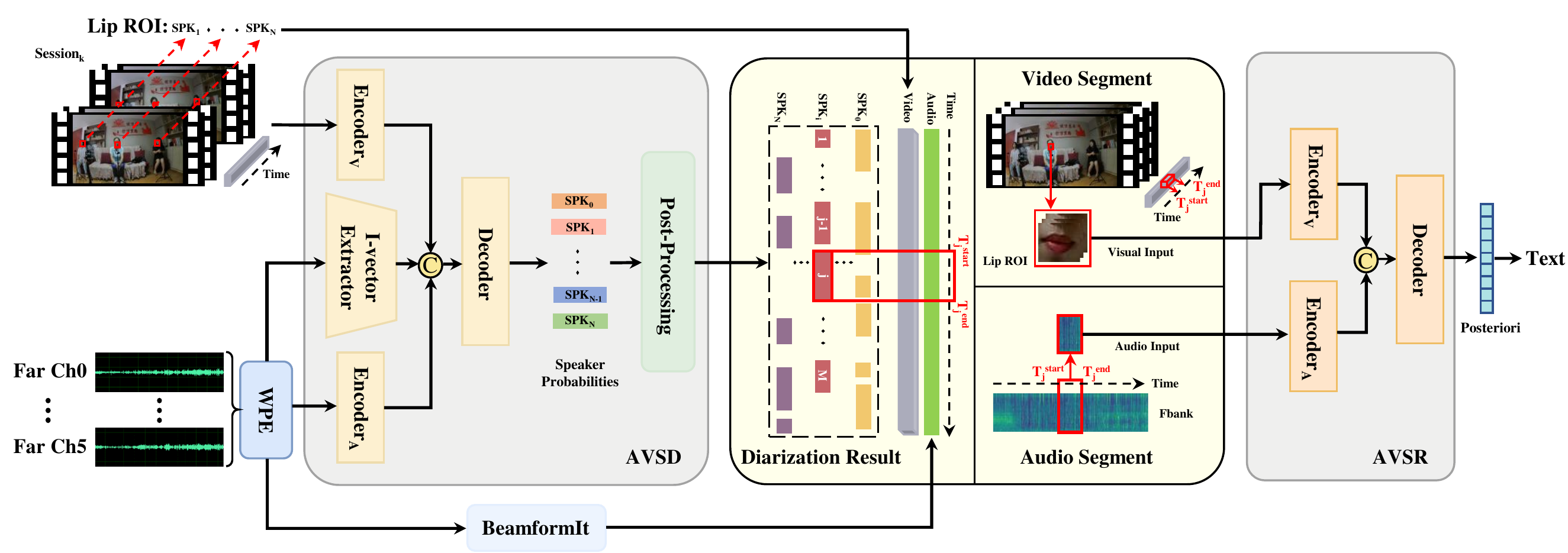}
\vspace{-0.65cm}
\caption{The architecture of the audio-visual diarization and recognition baseline system}
\label{baseline}
\vspace{-0.4cm}
\end{figure*}

\section{Dataset and Tracks}

\vspace{-0.15cm} 
\subsection{Training, Development, and Evaluation Sets}

We adopted the same training set as in the updated AVSR corpus of the MISP2021 challenge \cite{2022misptask2} and picked a new 3h development set from the previous development and evaluation sets. The new development set consists of the audio and video recordings of 8 rooms and 26 participants, including 10 males and 16 females. In the future, to eliminate overlapping speakers in each subset, a new evaluation set will be released and used for final ranking. The evaluation set only contains the recordings from the far-field devices.

\vspace{-0.15cm}
\subsection{Track 1: Audio-Visual Speaker Diarization}

Audio-visual speaker diarization aims to solve the ``who spoke when'' problem by labeling speech timestamps with classes that correspond to speaker identity using audio and video data. For evaluation, only the far-field audio and video data is available. We will provide the oracle speech segmentation timestamp. Participants need to submit a rich transcription time marked (RTTM) file for each session. RTTM files are text files containing one turn per line \cite{dihard}. The start time (4th column), duration (5th column), and speaker ID (8th column) must remain in the same columns.

Diarization error rate (DER) \cite{fiscus2006rich} is adopted as the evaluation metric. The lower the DER value (with 0 being a perfect score), the higher the ranking. It is worth noting that we do not set the ``no score'' collar, and overlapping speech will be evaluated. 
\begin{equation}
\rm DER \  =\   \frac{FA+MISS+SPKERR}{TOTAL}
\end{equation}
where FA, MISS, SPKERR are the total durations of the false alarm, missed detection and speaker error, respectively, and TOTAL is the sum of durations of all reference speakers' speech.

In Track 1, external audio data can be used to train the AVSD model, such as VoxCeleb 1, 2 \cite{nagrani2017voxceleb,chung2018voxceleb2}, CN-Celeb \cite{fan2020cn}, and other public datasets. Additional video data is also allowed to be used. However, participants should inform the organizers in advance about such data sources, so that all competitors know about them and have an equal opportunity to use them. 

\vspace{-0.15cm}   

\subsection{Track 2: Audio-Visual Diarization and Recognition}

Track 2 moves beyond AVSD and also considers the task of speech recognition, i.e., transcribing the speech into its verbatim text. The same evaluation set is adopted as Track 1. Participants need to submit the RTTM file, and transcription files. In each session, participants should chronologically merge all utterances from one speaker and provide a transcription file. Transcription files contain two columns: the utterance ID (1st column), and the utterance (2nd column). The format of the utterance ID is $ < \rm speaker \ ID>$\_$<\rm session \ ID>$.

With reference to the concatenated minimum-permutation word error rate (cpWER) in \cite{chime6}, we use concatenated minimum-permutation character error rate (cpCER) as the evaluation metric in Track 2. The calculation of cpCER is divided into three steps. First, recognition results and reference transcriptions belonging to the same speaker are concatenated on the timeline in a session. Second, character error rate (CER) of permutations of speakers is calculated as follows:
\begin{equation}
\rm CER \  =\   \frac{S+D+I}{N}
\end{equation}
where S, D, I are the character number of the substitution error, deletion error, and insertion error. N is the total number of characters. Finally, select the lowest CER as the cpCER.

In Track 2, we restrict the rules of additional data usage. External audio data and video data are allowed to be used. Significantly, participants can utilize timestamps, speaker tags, and other information except for text contents. Participants should also inform the organizers in advance about such data sources.

\vspace{-0.10cm} 
\section{BASELINE AVDR SYSTEM}
\label{sec:baseline}

Fig. \ref{baseline} shows the baseline AVDR system, which consists of an AVSD module followed by an AVSR module. The AVSD module also serves as the baseline system for Track 1. In this section, we elaborate the architecture and training process of the AVSD and AVSR modules, and provide the details about joining the AVSD and AVSR modules for decoding.

\vspace{-0.15cm}    
\subsection{Architecture and Training of the AVSD Module}
\label{ssec:avsdbase}

We follow our previous work \cite{avsd} as our baseline. The difference is that the preceding work used the data from the mid-field audio and video, while the current challenge focuses on the far-field audio and video.

As shown in the AVSD module in Fig. \ref{baseline}, our system has three encoder modules. In the visual encoder module, lip ROIs are used as input of the network which consists of lipreading model \cite{TCN}, conformer blocks \cite{conformer}, and a BLSTM layer. The whole network can be regarded as a visual voice activity detection (V-VAD) model to generate visual embeddings and an initial diarization result. Next, we use the audio dereverberated by NARA-WPE \cite{wpe} and the diarization result from the V-VAD model to compute i-vectors as speaker embeddings. Besides, through an FBank feature extractor and several 2D CNN layers, audio embeddings can also be extracted. In the decoder block, three types of embeddings are combined first and several BLSTM with projection (BLSTMP) layers are utilized to further extract features and get speech or non-speech probabilities for each speaker. In the post-processing stage, we ﬁrst perform thresholding with the probabilities to produce a preliminary result and adopt the same approaches as in \cite{he2022ustc}. Furthermore, DOVER-Lap \cite{dover-lap} is used to fuse the results of 6-channels audio.

The training process is as follows: first, we use the parameters of the pre-trained lipreading and train the V-VAD model. Then, we freeze the visual network parameters and train the audio network and decoder block. Finally, we unfreeze the visual network parameters, and train the whole network jointly.

\begin{table}
\caption{\textbf{Speaker diarization results on Dev set (in \%)}}
\centering
\setlength{\tabcolsep}{4.3mm}
\begin{tabular}{c|ccc|c}
\toprule
System &FA &\;MISS &SPKERR &DER\\
\midrule 
ASD &0.01&\;\;19.88&11.36&31.25 \\
VSD &6.64&\;\;8.17&3.89&18.69 \\
\textbf{AVSD}&\textbf{4.01}&\;\;\textbf{5.86}&\textbf{3.22}&\textbf{13.09} \\
\bottomrule
\end{tabular}
\vspace{-0.55cm}
\end{table}

\vspace{-0.15cm}   

\subsection{Architecture and Training of AVSR Module}
\label{ssec:avsrbase}
The AVSR model adopts a DNN-HMM hybrid system \cite{2022misptask2}. Firstly we apply the NARA-WPE \cite{wpe} and BeamformIt \cite{BF} to the far-field 6-channel speech. Then, the FBank features extracted from the audio and the Lip ROIs cropped from the video were segmented on the basis of the speaker diarization results. The front-end module composed of 3D convolution and ResNet-18 is used to extract lip-movement information for the video modality and outputs $ \rm embedding_V$. Meanwhile, the front-end module composed of 1D convolution and ResNet-18 is used to extract audio features and obtain $ \rm embedding_A$. The audio-visual features, $ \rm embedding_{AV}$, are extracted by the multi-stage temporal convolutional network (MS-TCN) \cite{ms-tcn} modules. Next, the posterior probabilities are obtained by the other MS-TCN modules. Finally, text is decoded from the posterior probabilities by using GMM-HMM, 3-gram model and DaCiDian.

During the training stage, oracle speaker diarization results are used. Kaldi \cite{povey2011kaldi} is applied to train a GMM-HMM system on all far-field audio data. The training of the DNN-based acoustic model uses Cross Entropy loss and Adam optimizer for 100 epochs with initial learning rate of 0.0003 and cosine scheduler. More details of the experiment can be found in \cite{2022misptask2}.

\vspace{-0.15cm}   

\subsection{Joint Decoding}
During inference, the RTTM file as the output of the  AVSD module contains the information of the $\rm Session, \rm SPK, \rm T^{start}$, and $\rm T^{dur}$. This information can be used for calculating DER in Track 1 and preprocessing far-field video and far-field 6-channel audio in Track 2. 
For $\rm Session_k$, a set of utterance identifier $(\rm SPK_i, \rm T_j^{start},\rm T_j^{dur})$ are available, where $\rm Session_k$ and $\rm SPK_i$ denote $k$-th session and $i$-th speaker in this session, $\rm T_j^{start}$ and $\rm T_j^{dur}$ denote the start time and the duration of the $j$-th utterance for $\rm SPK_i$. For the far-field video in $\rm Session_k$, we first segment the whole video according to $\rm T_j^{start}$ and $\rm T_j^{dur}$ and crop the lip region of $\rm SPK_i$ in every frame as the visual input of the AVSR module. For the far-field 6-channel audio in $\rm Session_k$, we first perform WPE and BeamformIt for the raw 6-channel audio and segment the whole beamformed audio according to $\rm T_j^{start}$ and $\rm T_j^{dur}$ as audio input of the AVSR module. Finally, we concatenate the decoded text of each utterance belonging to $\rm SPK_i$ in $\rm Session_k$ according to time order.

During the evaluation, due to the problem of permutation invariant training (PIT) and annotated segment text correspondence, we adopt cpCER as the final evaluation index.

\begin{table}
\caption{\textbf{Diarization and recognition results on Dev set (in \%)}}
\centering
\setlength{\tabcolsep}{3.65mm}
\begin{tabular}{c|ccc|c}
\toprule
System &S&D&I&cpCER \\
\midrule
ASR(OS)&40.84&27.33&0.51&68.68 \\
AVSR(OS)&35.78&27.82&0.36&63.96 \\
\midrule 
ASD+ASR&31.83&44.34&4.27&80.44 \\
VSD+ASR&39.25&31.22&0.66&71.13 \\
VSD+AVSR&35.17&31.01&0.61&66.79 \\
\textbf{AVSD+AVSR}&\textbf{35.94}&\textbf{29.45}&\textbf{0.68}&\textbf{66.07} \\
\bottomrule
\end{tabular}
\vspace{-0.45cm}
\end{table}

\vspace{-0.10cm} 
\section{Results AND ANALYSIS}
\label{sec:exp}
In this section, we first introduce the experimental results of the baseline systems. Next, we point out the difficulties of the MISP2022 challenge by providing examples, and analyze the good performance of AVDR system. Challenge participants can use this information to particularly focus on solving these issues in order to improve performance above the baseline.

\vspace{-0.15cm}   

\subsection{Baseline Results}

\label{ssec:results}
Table 1 shows the false alarm (FA) rate, missed detection (MISS) rate, speaker error (SPKERR) rate, and the DER for the audio-only speaker diarization systems (ASD), the visual-only speaker diarization system (VSD) and the audio-visual speaker diarization system (AVSD), where the latter is the baseline system of the AVSD track. For the ASD system, we use the VBx method \cite{VBx}. For the VSD system, we use the result from the visual encoder module as described in Section 3.1. The ASD system has poor results, most likely due to the loud TV background noises and high speaker overlap ratios, resulting in high MISS and SPKERR rates. Because the visual modality is not disturbed by the acoustic environment, the VSD system outperforms the ASD system in terms of MISS, SPKERR, and DER. However, VSD system has a high FA rate, potentially due to the lip movement in the silent segments. Combining the audio and visual modalities in the AVSD system yields the best performance, showing that both modalities can be combined to overcome their individual weaknesses.

\begin{figure}[ht]
\centering
\includegraphics[width=0.49\textwidth]{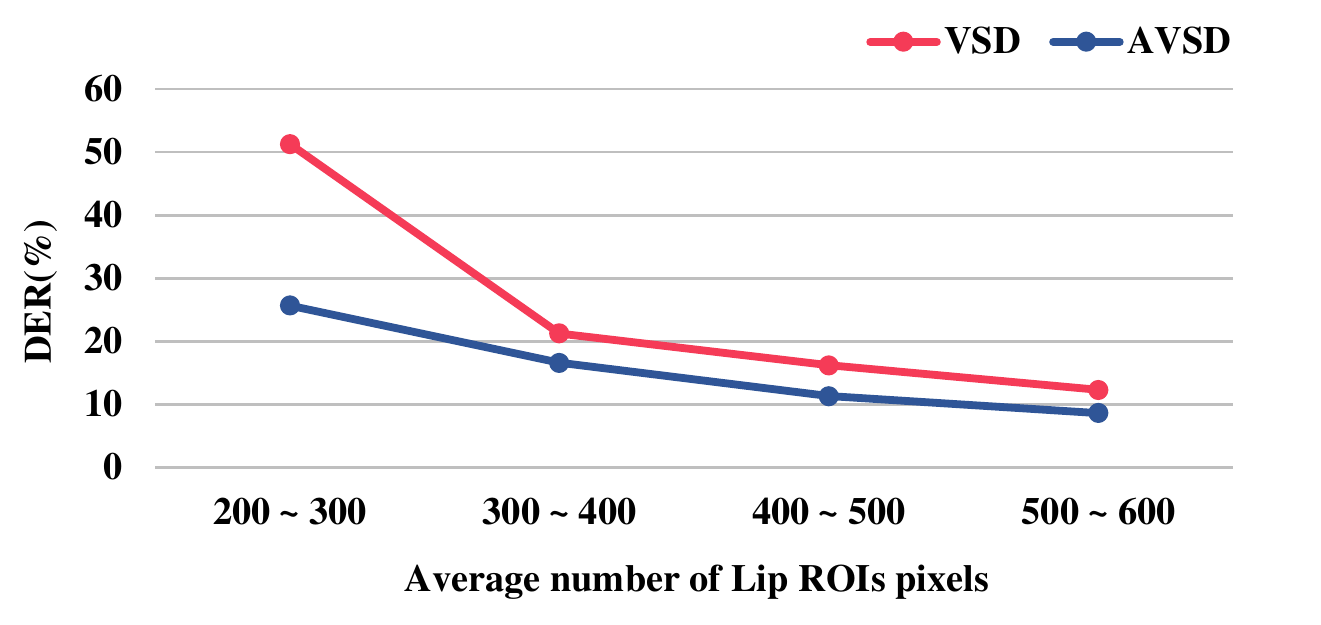}
\vspace{-0.85cm}
\caption{The DER comparison between the VSD and AVSD systems for different pixel values of Lip ROIS in the conversations}
\label{lip roi}
\vspace{-0.25cm}
\end{figure}

\begin{figure}[ht]
\centering
\includegraphics[width=0.49\textwidth]{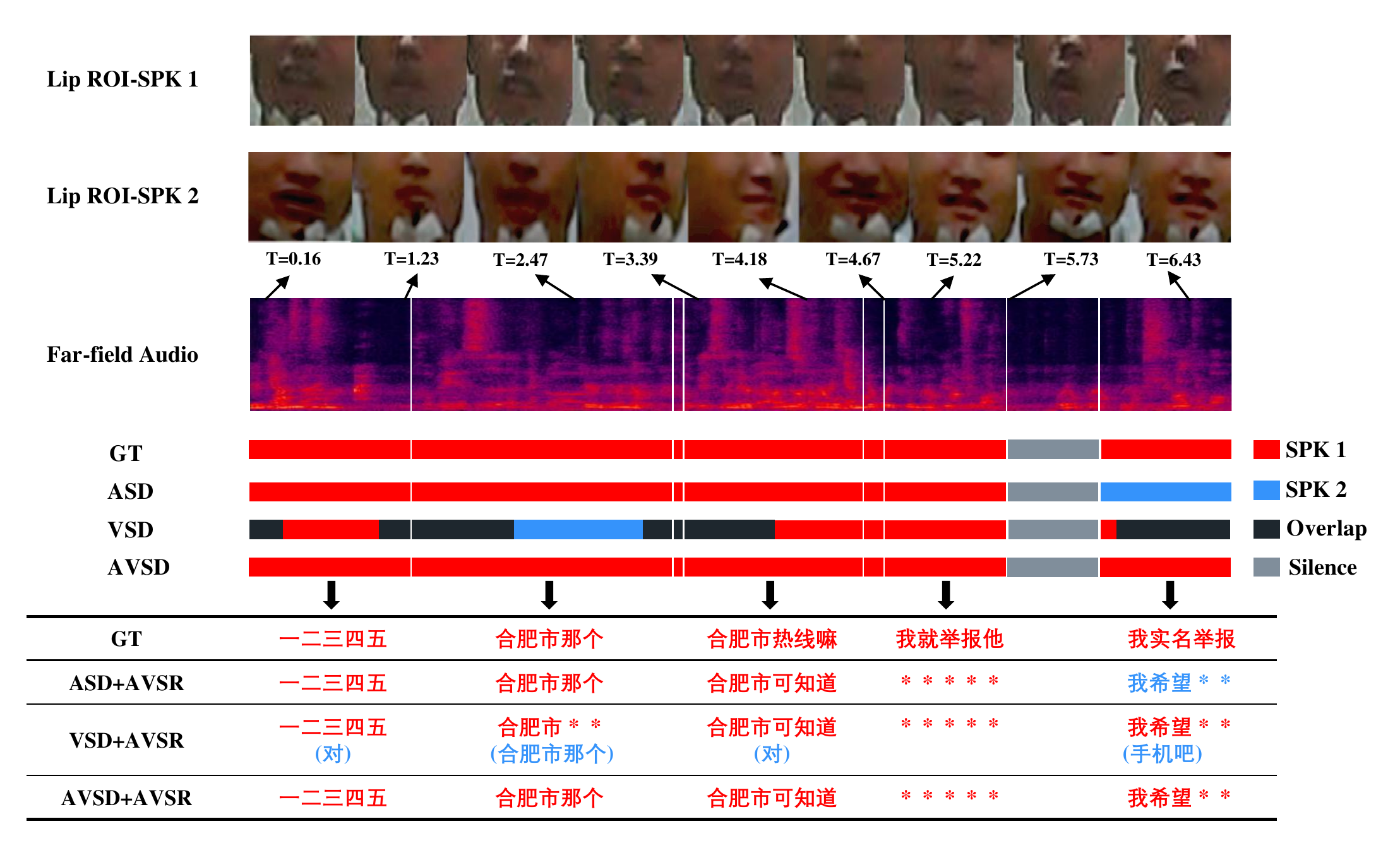}
\vspace{-0.65cm}
\caption{An example in a session with the comparison of results from different systems}
\label{example 1}
\vspace{-0.55cm}
\end{figure}

As shown in Table 2, we design 6 experiments for diarization and recognition system. The first two experiments are the speech recognition modules with the oracle speaker (OS) diarization results. The other experiments are the combinations of speaker diarization module and speech recognition module, e.g., ASD+ASR, VSD+ASR, VSD+AVSR, and AVSD+AVSR, where the latter is the baseline system of the AVDR track. For the ASD+ASR system, the high MISS and SPKERR rate results in a large number of deletion errors of target speakers. In addition, the high SPKERR rate leads to insertion errors of interfering speakers. Comparing the ASD+ASR system and the VSD+ASR system indicates that visual modality of speaker diarization module dominates the performance of the whole diarization and recognition system. In contrast to the VSD+ASR system, the visual modality in speech recognition module of the VSD+AVSR system provides distinguishable information that reduces substitution errors, which improves the whole system performance. In all experiments, it is the combination of the audio and visual modalities in both modules that yields the best system: AVDR.

\vspace{-0.15cm}   

\subsection{Analyses of difficulties}
\label{ssec:Analysis}
In order to let challenge participants solve problems better, we point out the potential difficulties in this challenge. Meanwhile, we discuss the performance of different module combinations to further explore the impact of audio and visual modalities.

\vspace{-0.20cm}

\subsubsection{Far-field Video Quality}
\label{sssec:Far-field video quality}
\vspace{-0.05cm}
Due to the long distance between cameras and speakers, far-field video will result in a greatly reduced proportion of each speaker's lip ROI in the total image, especially in the scenes with more speakers. At the same time, lamplight, position, angle, occlusion, and other environmental factors may lead to the reduction of video quality. We explore how the number of lip ROIs pixels affects the performance of the VSD and AVSD systems, as shown in Fig. \ref{lip roi}. It is found that as the average number of pixels decreases, the DER rises sharply. This will also affect the subsequent speech recognition task. 

\begin{figure}[ht]
\centering
\includegraphics[width=0.49\textwidth]{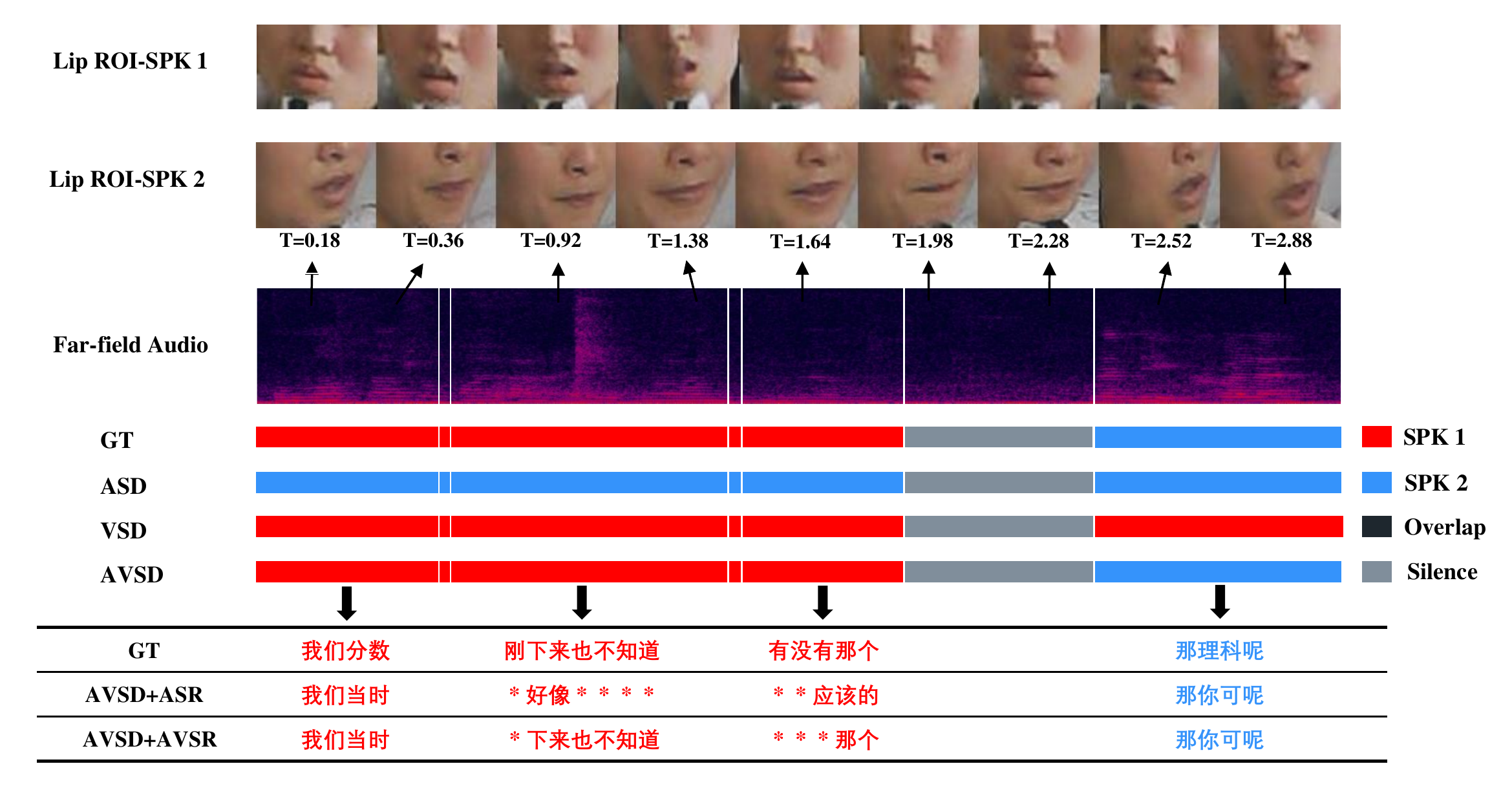}
\vspace{-0.65cm}
\caption{Another example in a session with the comparison of results from different systems}
\label{example 2}
\vspace{-0.55cm}
\end{figure}

According to the example in Fig. \ref{example 1}, it can be seen that dim-light and far distance lead to low quality of the far-field lip ROIs, making lip movements detection wrong or missing. There are lots of overlapping segment false detections and speaker confusion in the VSD results. In fact, according to the ground truth (GT), only one speaker (SPK 1) is talking all the time. For the module of AVSR using VSD results, the existence of overlapping segments leads to more insertion errors for SPK 2, and the speaker confusion leads to deletion errors for SPK 1. Because AVSD incorporates audio modality information to modify video modality, the results are significantly improved compared with VSD, and AVDR system results are also improved.

\vspace{-0.20cm}
\subsubsection{TV Background Noise}
\label{sssec:TV background noise}
\vspace{-0.05cm}
Since the TV is closer to the far-field microphone array, loud TV noise may cover the voice of the target speakers in the far-field audio. At the same time, due to the diversity of TV broadcast content, the audio may contain the voice of irrelevant speakers, which may interfere with the speaker diarization, and speech recognition. As shown in Fig. \ref{example 1}, in the fourth segment utterance, because actors on TV are talking loudly, noise received by the microphone completely covers the voice of the target speaker, making the AVDR system unable to recognize the speech content. Besides, in the last segment, due to the influence of TV background noise, ASD system wrongly assigns the segment of SPK1 to SPK2. Although the effect of AVDR system is better than that of single mode system, the TV background noise is also a big challenge in MISP2022.

\vspace{-0.20cm}
\subsubsection{Indistinguishable Speakers}
\label{sssec:spker}
\vspace{-0.05cm}
Due to the diversity of speakers, it is possible that speakers with similar timbre appear in the same session. As shown in Fig. \ref{example 2}, the similar timbre leads to speaker confusion in ASD result. In addition, peristalsis of lips, namely lip-movement without utterance, occasionally occurs in video recordings. It is difficult for the model to distinguish whether a speaker is talking or just moving his lip. In the VSD process, due to peristalsis, speaker confusion also arises which causes the target speaker to have more deletion errors and the interfering speaker to have more insertion errors. However, in the AVSD process, through the information complementation between audio-visual modalities, we get the diarization result consistent with the ground truth, which corrects the speech recognition errors caused by the wrong diarization result. 

\vspace{-0.20cm} 
\section{CONCLUSIONS}
\label{sec:conclusions}
\vspace{-0.10cm} 

This paper describes the MISP2022 challenge, which is the first to propose the audio-visual diarization and recognition (AVDR) task. We provide the analysis of this challenge, including the baseline results, the relationship between the diarization and the speech recognition modules, and the difficulties of the challenge. We believe that the research on audio-visual diarization and recognition can be better promoted through the MISP dataset and the MISP2022 challenge.

\vspace{-0.20cm} 
\section{ACKNOWLEDGEMENTS}
\vspace{-0.10cm} 
This work was supported by the National Natural Science Foundation of China under Grant No. 62171427.

\clearpage

\bibliographystyle{IEEEbib}
\bibliography{refs}

\end{document}